\documentclass[reprint,showpacs,amsmath,amssymb]{revtex4-1}

\usepackage[utf8]{inputenc}
\usepackage[T1]{fontenc}

\usepackage{amsmath}
\usepackage{mathrsfs}
\usepackage{txfonts}
\usepackage{mathtools}
\usepackage{braket}
\usepackage{tensor}
\usepackage{xcolor}
\usepackage[abbreviations]{siunitx}
\usepackage{graphicx}
\usepackage{booktabs}

\usepackage{float}

\usepackage{braket}

\usepackage{hyperref}
\usepackage{cleveref}

\newcommand{\beq}{\begin{equation}}
\newcommand{\eeq}{\end{equation}}
\newcommand{\beqn}{\begin{eqnarray}}
\newcommand{\eeqn}{\end{eqnarray}}
\newcommand{\bsub}{\begin{subequations}}
\newcommand{\esub}{\end{subequations}}
\newcommand{\bpm}{\begin{pmatrix}}
\newcommand{\epm}{\end{pmatrix}}

\begin{document}
    
\title{In-medium similarity renormalization group for a pairing-plus-particle-hole model }

 \author{L. H. Chen}  

 \affiliation{School of Physics and Astronomy, Sun Yat-sen University, Zhuhai 519082, P.R. China}

 \author{Y. G. Yao} 
 \affiliation{Okemos High School, Okemos,  MI 48864, USA}

  \author{B. C. He} 
 \email{Contact author: bhe9@utk.edu}
\affiliation{Department of Physics and Astronomy, University of Tennessee, Knoxville, TN 37996, USA}

\date{\today}

\begin{abstract} 

We benchmark two implementations of the in-medium similarity renormalization group (IMSRG) method, IMSRG(2) and IMSRG(2*), for the low-lying states of a pairing-plus-particle-hole model with varying numbers of fermions. In IMSRG(2), all operators are truncated up to the normal-ordered two-body terms, whereas IMSRG(2*) includes an additional term to partially account for higher-body contributions. The results are compared against exact solutions. We find that IMSRG(2*) consistently outperforms IMSRG(2) for both ground and excited states, although achieving convergence for excited states remains more challenging in strongly correlated systems than for the ground state.

\end{abstract}

\maketitle


\section{Introduction}
The quantum many-body problem remains one of the central challenges in physics due to its exponential complexity with increasing particle number and degrees of freedom. In nuclear physics, the past decade has witnessed remarkable progress in the development of ab initio methods for atomic nuclei~\cite{Hergert:2020abreview}, which solve the nuclear many-body problem using nonperturbative and systematically improvable techniques. Among these, the in-medium similarity renormalization group (IMSRG) method~\cite{Tsukiyama:2011PRL} has emerged as a particularly promising approach, with successful extensions to heavy nuclei up to $^{208}$Pb~\cite{Stroberg:2021PRL,Hu:2022NP} and deformed nuclei~\cite{Yao:2020PRL,Belley:2024PRL}. The core idea of IMSRG is to apply a sequence of continuous unitary transformations that gradually decouple a chosen reference state from the rest of the Hilbert space. This procedure partially diagonalizes the Hamiltonian, effectively evolving the reference state toward the ground state of the transformed Hamiltonian. As a result, the computational cost of IMSRG—both in memory and runtime—scales polynomially with the size of the single-particle basis and depends only indirectly on the number of particles~\cite{Hergert:2016PR}.

In practical applications of the IMSRG method, the normal-ordered two-body (NO2B) approximation is commonly used, where operators are truncated at the two-body level to balance accuracy and computational cost. This implementation, known as IMSRG(2) \cite{Hergert:2016PR}, becomes less reliable as interaction strength or nucleon number increases due to neglected induced higher-body terms. To address this, IMSRG(2) has been extended to IMSRG(3) \cite{Heinz:2021,Stroberg:2024,He:2024}, which includes all three-body terms but comes with a steep computational cost of $\mathcal{O}(N^9)$, compared to $\mathcal{O}(N^6)$ for IMSRG(2), where $N$ is the dimension of the single-particle basis. This limits IMSRG(3) to small model spaces, often insufficient for convergence. To overcome this, truncated IMSRG(3) schemes have been proposed. Notably, IMSRG(3N7), which retains essential three-body terms at a reduced cost of $\mathcal{O}(N^7)$, significantly improves accuracy over IMSRG(2) \cite{Stroberg:2024,Heinz:2021}. Additionally, the IMSRG(2*) approach has been developed to approximate higher-order contributions within the NO2B framework using auxiliary operators, while maintaining the same $\mathcal{O}(N^6)$ scaling as IMSRG(2). Benchmarking against the Lipkin-Meshkov-Glick model and even-mass carbon isotopes with a chiral two-body interaction shows that IMSRG(2*) outperforms IMSRG(2) \cite{Stroberg:2024}.
 
It is worth noting that IMSRG studies have primarily focused on ground states, while excited states are typically addressed in conjunction with other many-body methods~\cite{Hergert:2020abreview}. This raises the question of whether the IMSRG can be directly applied to excited states in quantum many-body systems. As emphasized in Ref.\cite{Hjorth-Jensen:2017}, the choice of reference state plays a crucial role in determining which eigenstate the IMSRG flow converges to. Notably, for solvable pairing Hamiltonians in the weakly interacting regime, IMSRG(2) has been shown to accurately reproduce specific excited states. Furthermore, Ref.\cite{Davison:2023} demonstrated that employing an ensemble reference state can mitigate unitarity violations introduced by truncations and enhance the description of low-lying states. In this work, we systematically investigate low-lying states in a pairing-plus-particle-hole (P3H) model using both IMSRG(2) and the modified IMSRG(2*) approach, benchmarking their performance against exact solutions over a wide range of particle numbers, model space sizes and interaction strengths.

The article is organized as follows. In Sec.\ref{scetion:methods}, we briefly introduce the pairing-plus-particle-hole(P3H) model, the IMSRG(2) method and its extensions IMSRG(2*). The results are discussed in  Sec. \ref{section:result}. The conclusions of this study and outlooks are given in Sec.\ref{section:summary}

\section{The methods}
\label{scetion:methods}

\subsection{The P3H model}

\begin{figure}[bt]
\centering
\includegraphics[width=\columnwidth]{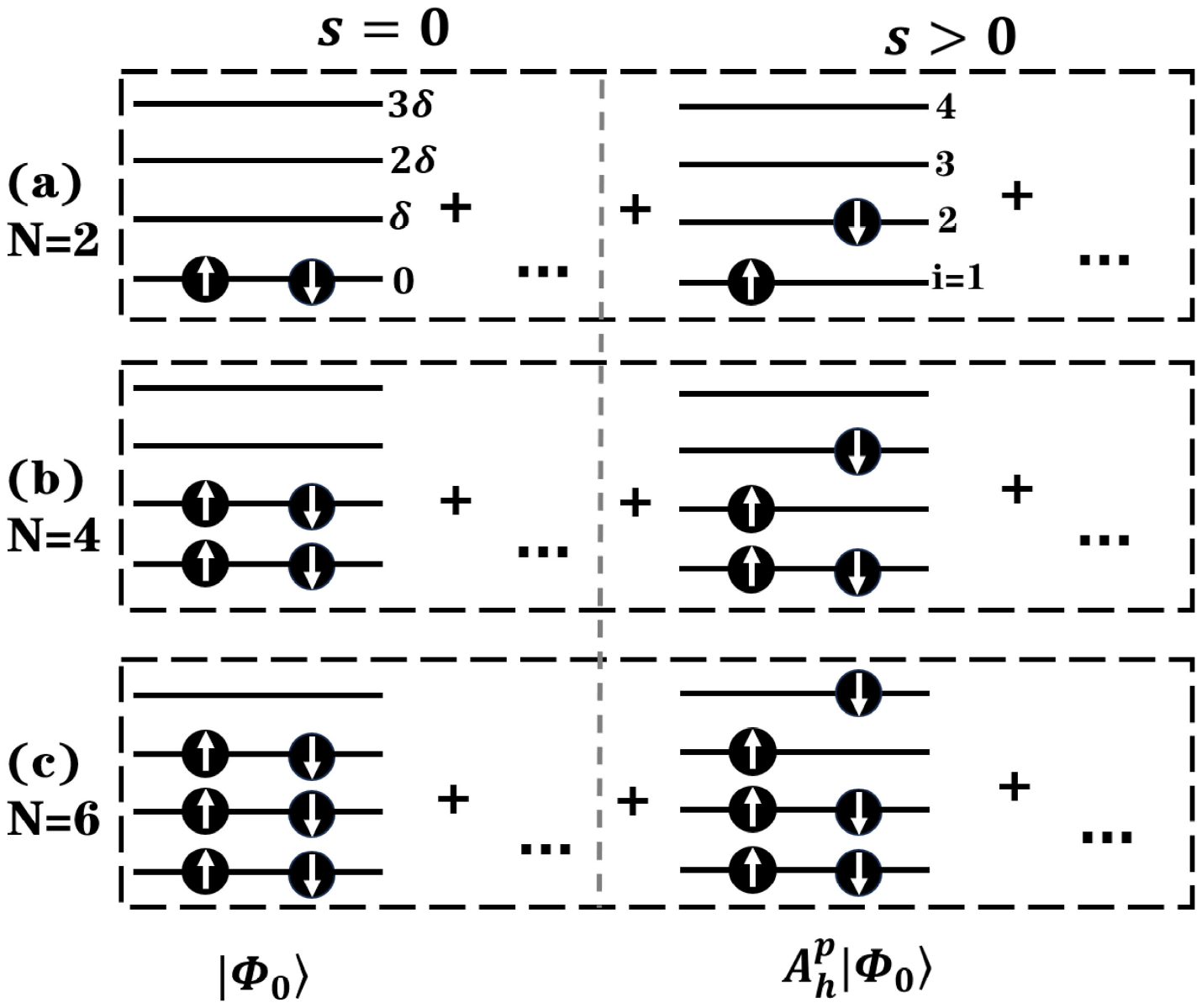}
\caption{A schematic picture for the configurations of the $N$-particle system  in the P3H model with particle number $N=2, 4$, and $6$, respectively. The configurations are classified into two types, i.e., fully occupied ones with the seniority number $s=0$ (left), and  the ones with $s >0$ (right).}
 \label{fig:pairing_model_figure}
 \end{figure}

The Hamiltonian of the P3H model is given by~\cite{Davison:2023}
\beqn
\label{eq:H}
    H_0  &=& \delta \sum_{i} (|i|-1) a_{i}^\dagger a_{i} - \frac{g}{2} \sum_{i,j>0} a_{i}^\dagger a_{\bar i}^\dagger a_{\bar j} a_{j} \nonumber\\ 
    &-&\frac{b}{2} \sum_{i,j,l>0} \left( a_{i}^\dagger a_{\bar i}^\dagger a_{\bar j} a_{l} + a_{l}^\dagger a_{\bar j}^\dagger a_{\bar i} a_{i} \right), 
\eeqn
where $\delta$ represents the energy gap between single-particle levels indexed by the principal quantum number $|i| = 1, 2, \cdots, L$, and $(i, \bar{i})$ denotes the $i$-th state with spin up and spin down, respectively. The parameter $g$ is for the strength of the pairing interaction, and $b$ characterizes the pairing-breaking interaction. In this work, we mainly consider the single-particle model space $L = 4$ and only focus on pairing correlations between particles with opposite spin directions, even though we will finally examine the impact of model space size on the performance of the methods. The total number of particles $N$ ranges from $N=2$ to $N=6$, as illustrated in Fig.~\ref{fig:pairing_model_figure}(a)–(c), corresponding to systems with valence particles, mid-shell occupancy, and valence holes, respectively, thereby simulating typical scenarios in valence-space shell-model calculations.

The  lowest-energy mean-field configurations for the  $N$-particle system can be written as
\beq 
\ket{\Phi_0 (N) } = \prod^{N}_{i}  a^\dagger_{i} \ket{0},
\eeq 
which can be conveniently rewritten in the occupation-number representation 
\begin{equation} 
    | \Phi_0 (N) \rangle  = \ket{n_1n_{\bar 1} \cdots n_in_{\bar i}\cdots}=|\underbrace{1111}_{N}\underbrace{0000\dots }_{2L-N}\rangle,
\end{equation}
 with $n_i=1$ representing fully occupied and $n_i=0$ representing unoccupied $i$-th state, respectively.  The excited configurations can be written as many-particle-many-hole (mp-mh) excitations on top of the $| \Phi_0 (N) \rangle$,
 \beqn 
 A^{pp'\cdots}_{hh'\cdots} | \Phi_0 (N) \rangle,\quad 
 A^{pp'\cdots}_{hh'\cdots} \equiv a^\dagger_p a^\dagger_{p'}\cdots a_{h'}a_h.
 \eeqn

In the configuration-interaction (CI) approach based on the diagonalization method, the wave function of the $k$-th eigenstate of the Hamiltonian is constructed as a linear combination of the above basis configurations
\begin{equation}
    \ket{\Psi_k} = \sum_{\alpha=0} c^{(k)}_\alpha \ket{\Phi_\alpha},
\end{equation}
where the expansion coefficient $c^{(k)}_\alpha$ is obtained from the diagonalization of the Hamiltonian matrix defined in the configuration space, see Fig.~\ref{fig:pairing_model_figure}. In this work, we consider only the configurations $\ket{\Phi_\alpha}$ with total spin projection $S_z = \sum_{i=1}^N s_z(i) = 0$. Under this constraint, the configurations $\ket{\Phi_\alpha}$ can be categorized into two types: (i) fully paired configurations $\ket{\Phi_\alpha}(s = 0)$, characterized by a seniority number $s = 0$; and (ii) pairing-broken configurations $\ket{\Phi_\alpha}(s > 0)$, where the seniority is nonzero.

\subsection{The IMSRG}
The basic idea of the IMSRG method is decoupling  the off-diagonal elements of a Hamiltonian in the configuration space by introducing a set of continuous unitary transformations~\cite{Hergert:2016PR} 
\begin{equation}
    H(s) \equiv U(s) H(0) U^\dagger(s).
\end{equation}
Here, $H(0)$ is the initial Hamiltonian. The IMSRG flow equation can be derived by differentiating the transformed Hamiltonian $H(s)$ with respect to the flow parameter $s$,
\begin{equation}
\label{eq:SRG_flow_equation}
    \frac{d}{ds} H(s) = [\eta(s), H(s)],
\end{equation}
where the generator $\eta(s)$ is defined by
\begin{equation}
    \eta(s) \equiv \frac{dU(s)}{ds} U^\dagger(s) = -\eta^\dagger(s). 
\end{equation}
In the practical application, the unitary transformation $U(s)$ is determined by the generator $\eta(s)$ instead. 
 
It is convenient to rewrite the Hamiltonian (\ref{eq:H}) in a general form
\begin{equation}
    H(0) = \sum_{ij} T^i_j  A^i_j + \frac{1}{4} \sum_{ijkl} V^{ij}_{kl} A^{ij}_{kl},
\end{equation}
where  $T^{i}_{j}= \delta_{ij} (|i|-1)\delta$, and the matrix element $V^{ij}_{kl}$ is  given by
\begin{equation}
    V^{ij}_{kl}  = 
\begin{cases}
-g/2 , &  \delta_{j\bar i}\delta_{l\bar k} \\
-b/2, 
&  \delta_{j\bar i}\Delta_{k\neq l}X_{kl}~{\rm or}~\delta_{l\bar k}\Delta_{i\neq j}X_{ij} \\
0 & \text{else}
\end{cases}.
\end{equation}
In the above matrix element, $X_{kl}=0$ if the  $k$ and $l$ states have the same spin direction, and $X_{kl}=1$ if they have opposite spin directions.
Moreover, $\Delta_{k\neq l}=1$ for $k\neq l$; otherwise, it is zero. 

It is worthy to note that the commutator on the right-hand side of the Eq.(\ref{eq:SRG_flow_equation}) can induce $N$-body operators. In the IMSRG(2),  all the operators are truncated up to normal-ordered two-body (NO2B) terms for simplicity.  For a given single-reference state $|\Phi_{\rm Ref} \rangle$, we normal-order the transformed Hamiltonian $H(s)$ as below
\begin{equation}
\label{eq:transformed_H}
    H(s) = E(s) + \sum_{ij} f_j^i(s) \{ A_j^i \} + \frac{1}{4} \sum_{ijkl} \Gamma_{kl}^{ij}(s) \{ A_{kl}^{ij} \},
\end{equation}
where $\{A_{lmn\ldots}^{ijk\ldots} \}$ denotes the normal-ordered operator, whose expectation value with respect to the reference state is zero. The  normal-ordered zero-body term $E$ represents the expectation value of the $H(0)$ with respect to the reference state,
\begin{equation}
\label{eq:NO-0B E} 
    E(0) =  \bra{\Phi_{\rm Ref}}  H(0)\ket{\Phi_{\rm Ref}},
\end{equation}
the normal-ordered one-body (NO1B) term,
\begin{equation}
\label{eq:NO-1B f}
    f^{i}_{j}(0) = T^{i}_{j} + \sum_k V^{ik}_{jk} n_k,
\end{equation}
and the two-body term
\begin{equation}
\label{eq:NO-2B Gamma}
    \Gamma^{ij}_{kl}(0) = V^{ij}_{kl},
\end{equation} 
where the occupation number is defined as the eigenvalue of the one-body density matrix on the reference state,
\beq
n_k = \bra{\Phi_{\rm Ref}} A^k_l \ket{\Phi_{\rm Ref}}\delta_{kl}.
\eeq 
For $s\neq0$, they are determined by the flow equation (\ref{eq:SRG_flow_equation}), which is transformed into a set of ordinary differential equations (ODEs),
\bsub
\beqn  
\label{eq:IMSRG(2)_dE}
    \frac{dE}{ds} &=& \sum_{ab} (n_a - n_b) \eta^{a}_{b} f^{b}_{a} + \frac{1}{2} \sum_{abcd} \eta^{ab}_{cd} \Gamma^{cd}_{ab} n_a n_b \bar{n}_c \bar{n}_d, \\
\label{eq:IMSRG(2) df} 
    \frac{df^{i}_{j}}{ds} &=& \sum_a (1 + P_{ij}) \eta^{i}_{a} f^{a}_{j} + \sum_{ab} (n_a - n_b) (\eta^{a}_{b} \Gamma^{bi}_{aj} - f^{a}_{b} \eta^{bi}_{aj}) \nonumber\\
&&+ \frac{1}{2} \sum_{abc} (n_a n_b \bar{n}_c + \bar{n}_a \bar{n}_b n_c) (1 + P_{ij}) \eta^{ci}_{ab} \Gamma^{ab}_{cj},\\
\label{eq:IMSRG(2) dGamma} 
    \frac{d\Gamma^{ij}_{kl}}{ds} 
    &=& \sum_a \left\{ (1 - P_{ij})(\eta^{i}_{a}\Gamma^{aj}_{kl} - f^{i}_{a}\eta^{aj}_{kl}) - (1 - P_{kl})(\eta^{a}_{k}\Gamma^{ij}_{al} - f^{a}_{k}\eta^{ij}_{al}) \right\} \nonumber\\
&&+ \frac{1}{2} \sum_{ab} (1 - n_a - n_b)(\eta^{ij}_{ab}\Gamma^{ab}_{kl} - \Gamma^{ij}_{ab}\eta^{ab}_{kl}) \nonumber\\
&&+ \sum_{ab} (n_a - n_b)(1 - P_{ij})(1 - P_{kl})\eta^{ai}_{bk}\Gamma^{bj}_{al}.
\eeqn
\esub

Here, $\bar{n}_i = 1- n_i$, and $P_{ij}$ is  the permutation operator that exchanges the indices  $i$ and $j$.  The $\eta^{i}_{j}$ and $\eta^{ij}_{kl}$
denote the matrix elements of  NO1B and NO2B parts of the generator $\eta$. In this work, we choose different types of generators. Considering the observation that the White generator is generally very efficient for well-behaved problems~\cite{Morris:2015}, we mainly employ the White generator~\cite{White2002}
\beqn
\label{eq:SRG_white_generator}
    \eta^{\text{Wh}}  
    =  \sum_{ph} { f^{p}_{h}(s)}{\Delta^{p}_{h}(s)} \{ A^{p}_{h} \} + \frac{1}{4} \sum_{pp'hh'}  \frac{\Gamma^{pp'}_{hh'}(s)}{\Delta^{pp'}_{hh'}(s)} \{ A^{pp'}_{hh'} \} - {\rm H.c.}, 
\eeqn  
where H.c. denotes the Hermitian conjugate. In the case of degenerate or near-degenerate energy levels leading to excessively small $\Delta^{p}_{h}$ or $\Delta^{pp'}_{hh'}$ values, we automatically switch to the arc-tangent variant of the White generator to handle the singularity problem, which is given by~\cite{White2002,Hergert:2017PS} 
\beqn
\label{eq:SRG_white_arctan_generator}
    \eta^{\text{Wh-atan}}  
    &=& \frac{1}{2} \sum_{ph} \arctan \frac{2 f^{p}_{h}(s)}{\Delta^{p}_{h}(s)} \{ A^{p}_{h} \} \nonumber\\ 
    &&+ \frac{1}{8} \sum_{pp'hh'} \arctan \frac{2 \Gamma^{pp'}_{hh'}(s)}{\Delta^{pp'}_{hh'}(s)} \{ A^{pp'}_{hh'} \} - {\rm H.c.}, 
\eeqn
In the case that both the White generator and its arc-tangent variant do not work, we choose the imaginary-time generator instead \cite{Hergert:2017PS},
\begin{equation}
\begin{aligned}
    \eta^{\rm IM}=&  \sum_{ph} \text{ sgn} (\Delta^{p}_{h}(s)) f^{p}_{h}(s) \{ A^{p}_{h} \} \\ 
    &+ \frac{1}{4} \sum_{pp'hh'} \text{sgn} ({\Delta^{pp'}_{hh'}(s)}) \Gamma^{pp'}_{hh'}(s) \{ A^{pp'}_{hh'} \} - {\rm H.c.},
\end{aligned}
\end{equation}
which turns out to improve the convergence behavior in certain cases, as demonstrated in this work.




In the above expressions, the Epstein–Nesbet matrix elements are defined as~\cite{Hergert:2017PS}.
\begin{equation}
    \Delta^{p}_{h} = f^{p}_{p}  - f^{h}_{h}  + \Gamma^{ph}_{ph},
\end{equation}
and
\beqn 
    \Delta^{pp'}_{hh'} 
    &=& f^{p}_{p} + f^{p'}_{p'} - f^{h}_{h} - f^{h'}_{h'} + \Gamma^{hh'}_{hh'} + \Gamma^{pp'}_{pp'} \nonumber\\
&&- \Gamma^{ph}_{ph} - \Gamma^{p'h'}_{p'h'} - \Gamma^{ph'}_{ph'} - \Gamma^{p'h}_{p'h}.
\eeqn

Following Ref.~\cite{Stroberg:2024}, we  implement the IMSRG(2*), in which  one incorporates an additional operator into the flow equation to simulate the contributions of higher-body terms while still truncating at the two-body level.  The Eq.(\ref{eq:SRG_flow_equation}) is modified as follows
\begin{equation}
    \frac{d}{ds}H(s) = [\eta(s), H(s) + \chi(s)],
\end{equation}
where $\chi(s)$ is an auxiliary NO1B operator, determined by the flow equation
\begin{equation}
    \frac{d}{ds} \chi_{ij} = (n_i n_j + \bar{n}_i \bar{n}_j) [\eta_{\rm 2B}, H_{\rm 2B}]_{ij}, 
\end{equation}
where $\eta_{\rm 2B}$ and $H_{\rm 2B}$ denote the two-body parts of the $\eta(s)$ and $H(s)$.

\section{Results and discussion}
\label{section:result}

\begin{figure}[bt]
\centering
\includegraphics[width=\columnwidth]{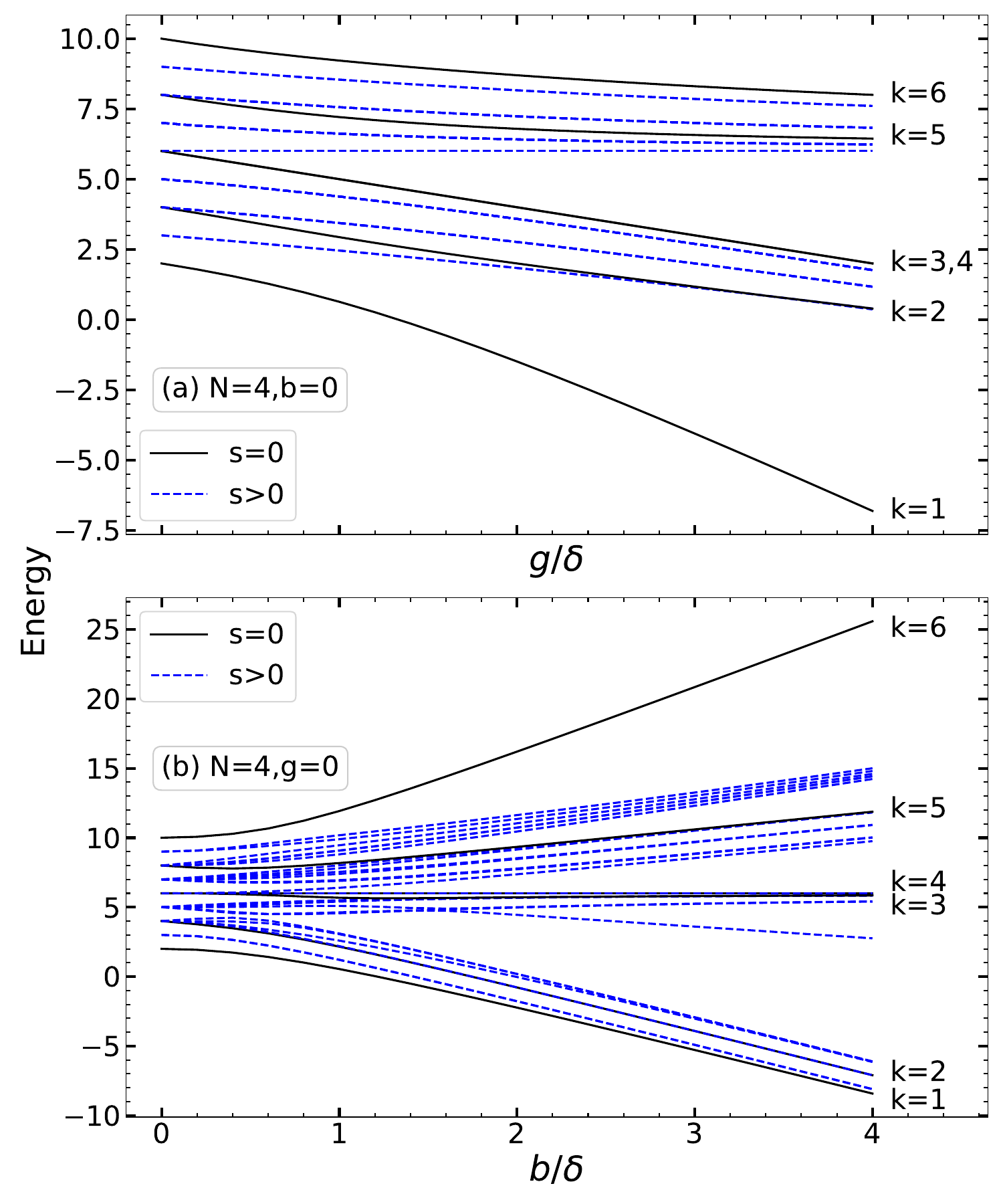}
\caption{(Color online) The energy levels of the P3H model as a function of (a) pairing interaction strength $g/\delta$ with $b=0$; (b) and pairing-breaking interaction strength $b/\delta$ with $g=0$. The black solid curves indicate the lowest-energy levels, which are split from the states with $b=0$(a) or $g=0$(b), built on the fully-paired configurations with the seniority number $s=0$. }
 \label{fig:energy_levels_exact_gb}
 \end{figure}

\begin{figure}[bt]
\centering
\includegraphics[width=\columnwidth]{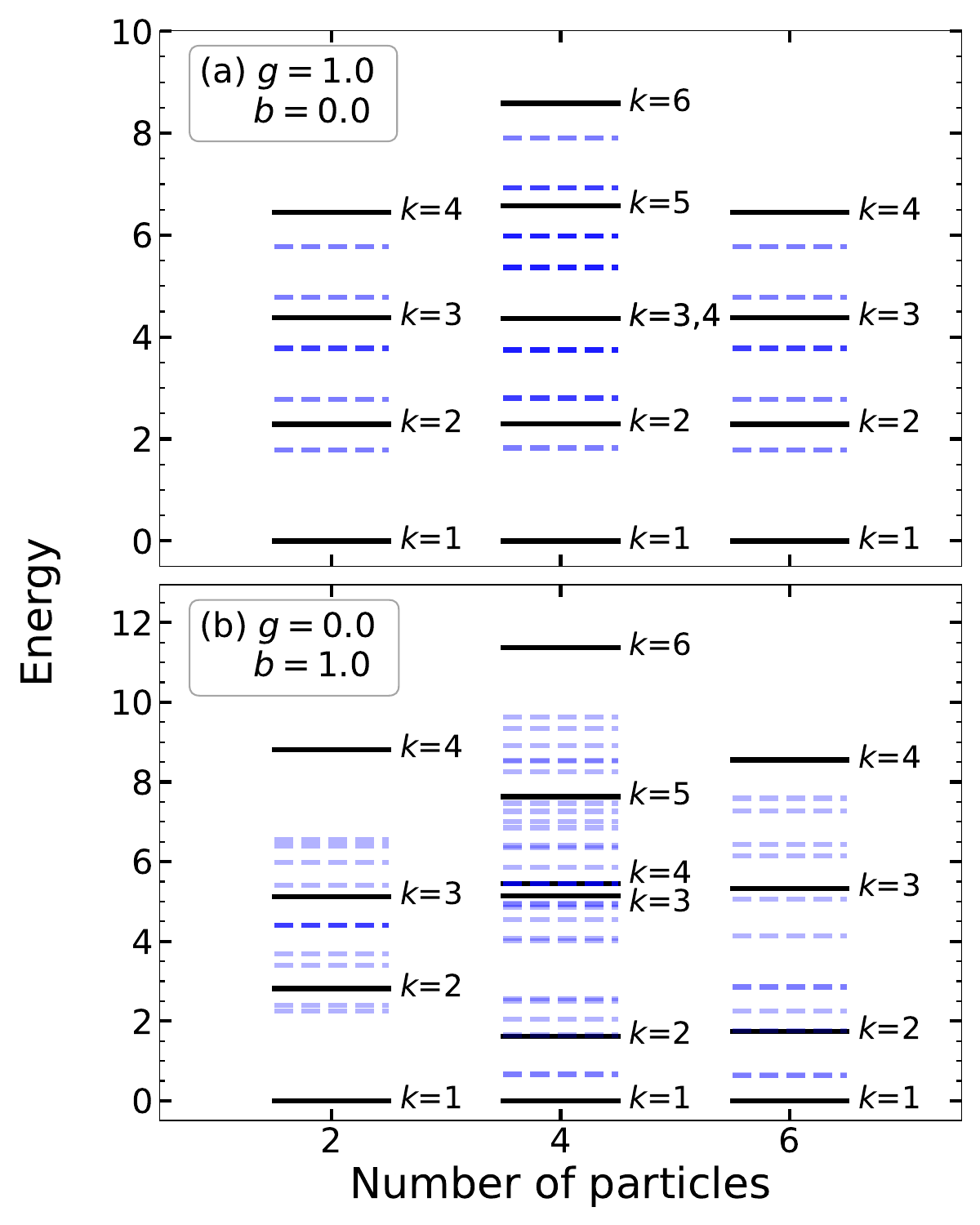}
\caption{(Color online) Same as Fig. \ref{fig:energy_levels_exact_gb}, but for the energy levels of the P3H model with $N=2, 4$ and $6$ particles using (a)  $g=1, b=0$; (b)   $g=0, b=1$.  All energy levels are normalized to the ground-state energy.}
 \label{fig:energy_levels_exact}
 \end{figure}

The energy spectra of the $N=4$ system from the exact solution of the P3H model for different interaction parameters are displayed in Fig.~\ref{fig:energy_levels_exact_gb}.  As shown in Fig.~\ref{fig:energy_levels_exact_gb}(a), for the $b = 0$ case, all energy levels are separated into two distinct categories: one constructed from fully paired configurations with the seniority number $s=0$, and the other from pairing-broken configurations with $s>0$. As expected, the ground-state energy decreases rapidly with the increase  of pairing strength $g$, resulting in an increasing excitation energy of the first excited state. For the pairing-broken case with strength parameter $b\neq0$, several energy levels are split with the increase of $b$ and all the states are admixtures of pairing and pairing-broken configurations. For convenience, we still use black solid curves to indicate the  lowest-energy levels split from the states at $b=0$  built on the fully-paired configurations, as these levels are the most probably reachable in the IMSRG calculation with the choice of a fully-paired reference state.

Figure \ref{fig:energy_levels_exact} compares the energy spectra of the  systems with $N=2, 4$ and $6$ from the exact solution for the P3H model with either $g=1, b=0$ or $g=0, b=1$. For the case of $b=0$, the energy spectra of $N=2$ and $N=6$ systems are exactly the same. In contrast, for the $b\neq0$ case, they are different as the pairing-breaking term in the Hamiltonian (\ref{eq:H}), characterized by the strength parameter $b$, violates the symmetry between particles and holes.

\begin{figure}[bt]
\centering
\includegraphics[width=\columnwidth]{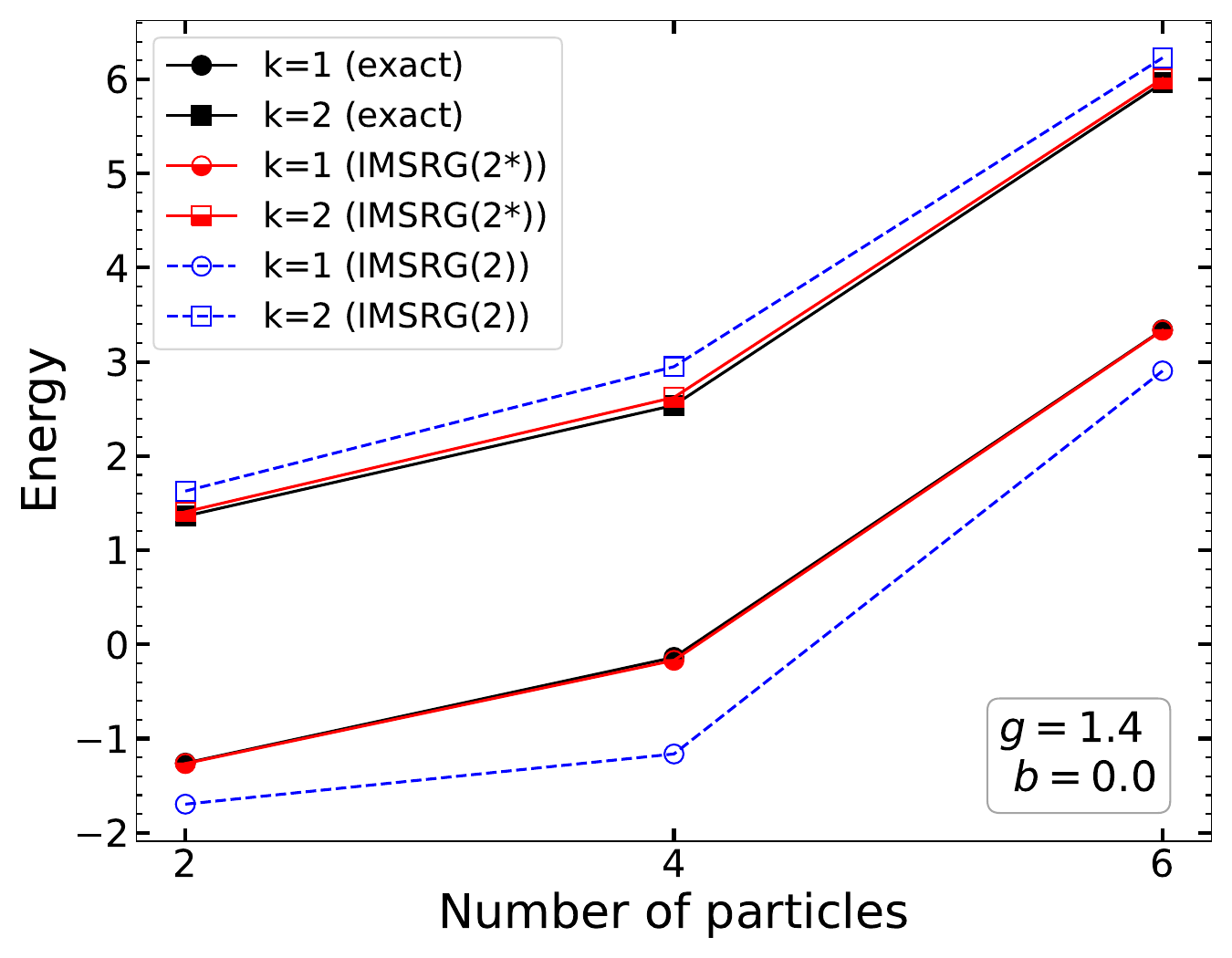}
\caption{(Color online) The energies of the $k=1, 2$ states by the IMSRG(2) and the IMSRG(2*) methods for  the $N=2, 4$ and $6$ using $g=1.4, b=0$.  }
\label{fig:Energies_different_N}
\end{figure}

\begin{figure}[bt]
\centering
\includegraphics[width=\columnwidth]{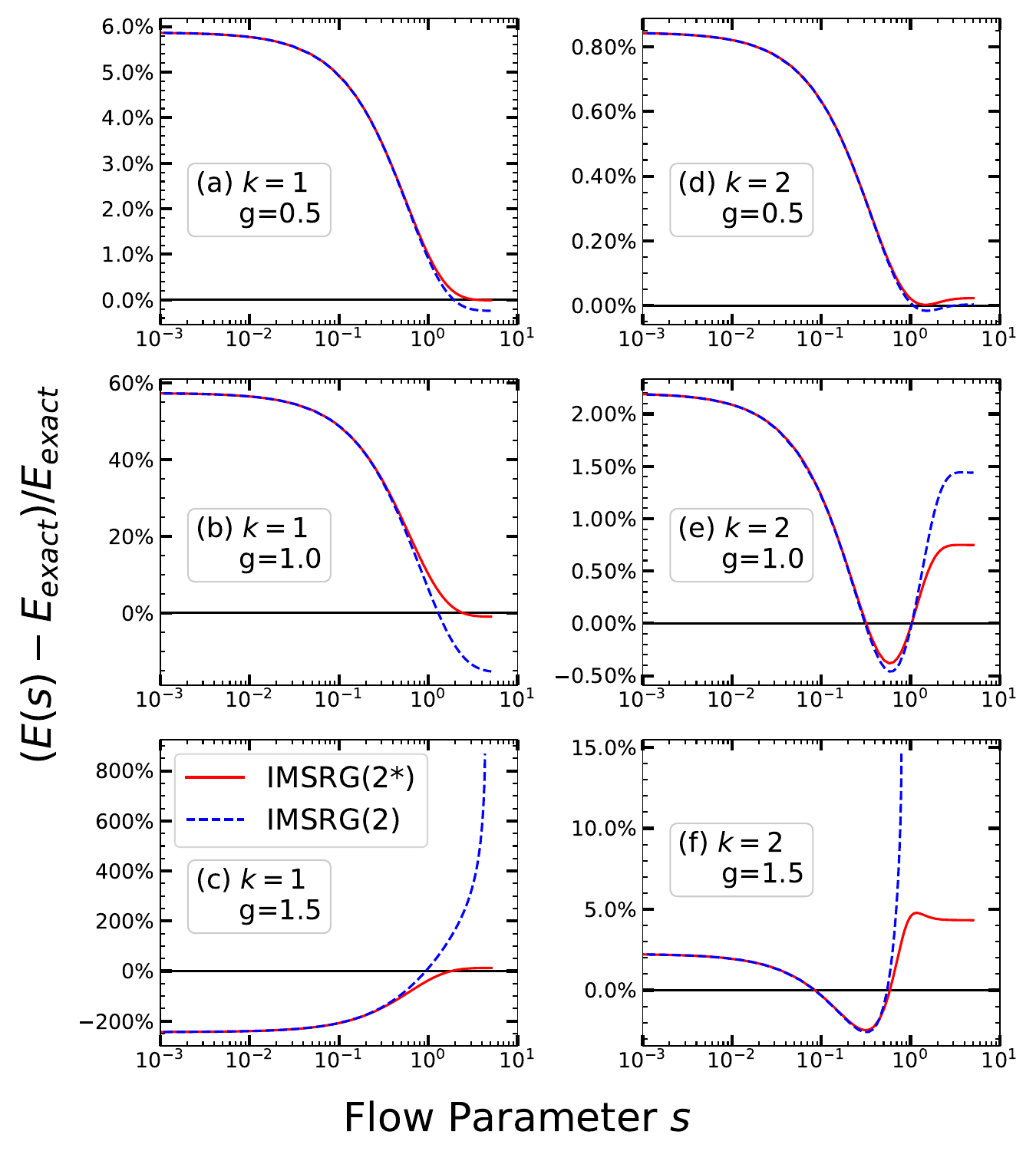}
\caption{(Color online) Comparison of the energy flows in the IMSRG(2* )(red) and IMSRG(2)(blue) methods for the $k=1, 2$ states  with the flow parameter $s$ for  $N =4$, $g =0.5$, $1.0$, $1.5$ and $b=0$, respectively.}
\label{fig:energy_flow}
\end{figure}

\begin{figure}[bt]
\centering
\includegraphics[width=\columnwidth]{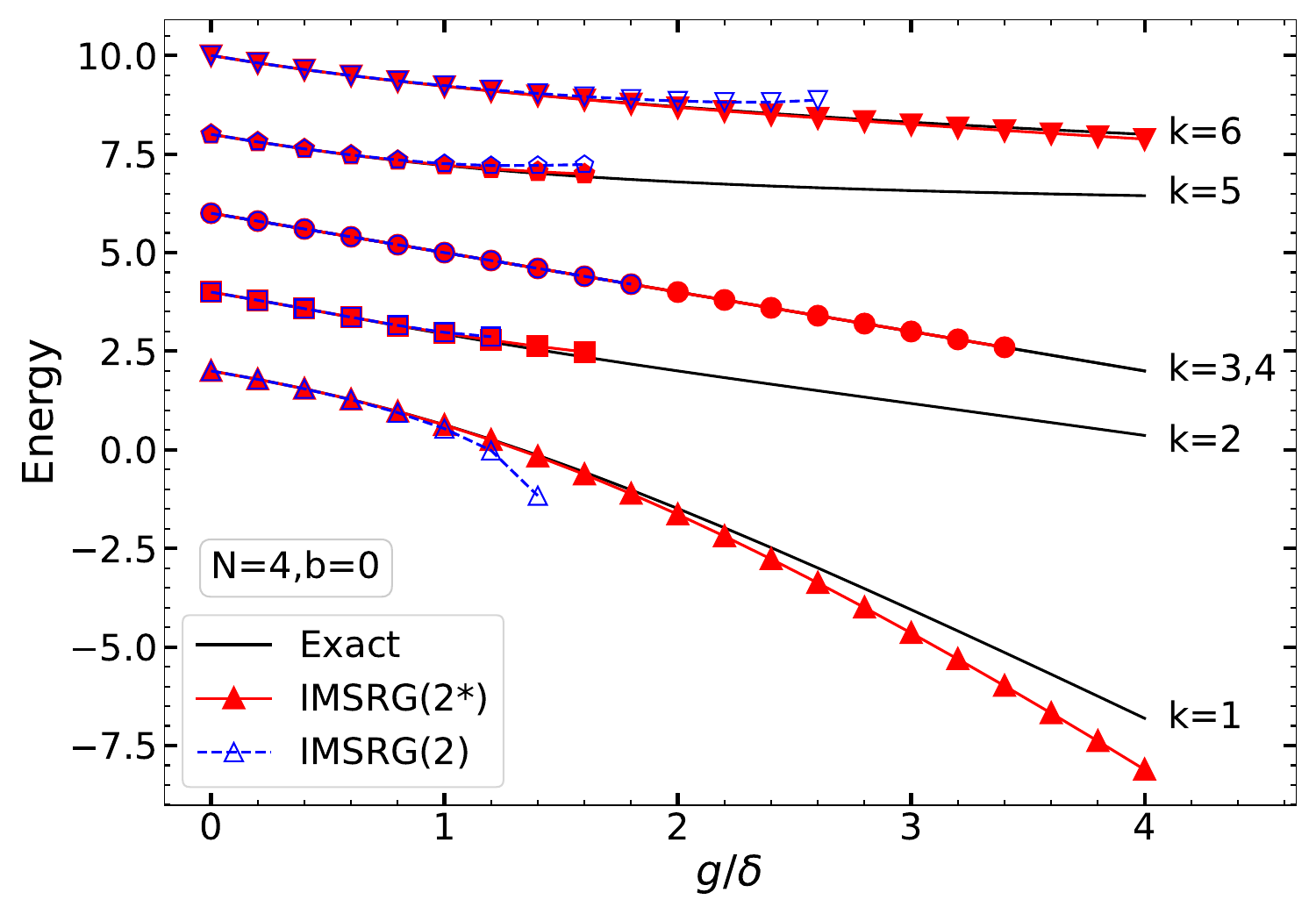}
\caption{(Color online)  Comparison of the energy levels by the  IMSRG(2), IMSRG(2*) and exact solutions for $N=4, b=0$, but with different pairing strength $g$.  }
\label{fig:N4_Levels6}
\end{figure}

Figure~\ref{fig:Energies_different_N} compares the energies of the $k = 1, 2$ states calculated using the IMSRG(2) and IMSRG(2*) methods for systems with $N = 2, 4, 6$, using $g = 1.4$ and $b = 0$, in comparison with the exact solutions. The discrepancy between IMSRG(2) and the exact results is noticeably larger for $N = 4$ than for $N = 2$ and $N = 6$. It can be understood from the fact that many-body correlation in the $N=4$ system is stronger than that in the $N=2, 6$ systems for the given model space with $L=4$, presenting a more challenge for the IMSRG(2) method. The discrepancy is mitigated significantly in the IMSRG(2*) method.
As shown in Fig.~\ref{fig:energy_flow}, the convergence behavior of the $k = 1, 2$ energy levels is also improved in IMSRG(2*). In particular, Fig.~\ref{fig:N4_Levels6} demonstrates a systematic improvement in the description of higher-lying excited states for various values of $g$ (with $b = 0$). Furthermore, IMSRG(2*) provides a reasonable description of the energy levels even in the regime of pairing strength $g$ where the IMSRG(2) flow fails.

\begin{figure}[bt]
\centering
\includegraphics[width=\columnwidth]{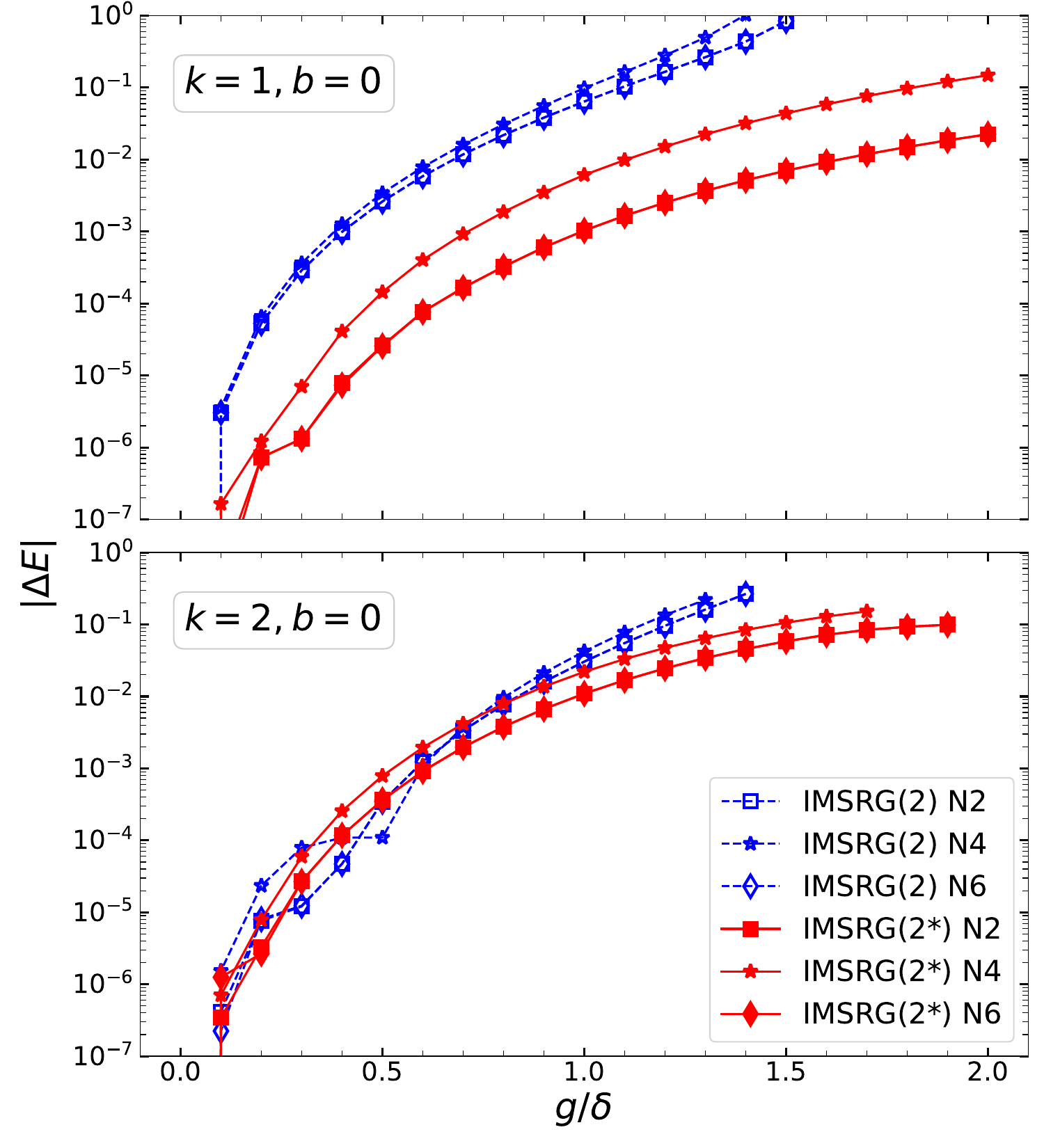}
\caption{(Color online)  The discrepancies $|\Delta E|$ in the energies of the $k = 1, 2$ states, obtained from the IMSRG(2) and IMSRG(2*) methods relative to the exact solutions, as functions of the pairing strength $g$ for systems with $N = 2, 4$, and $6$.}
\label{fig:discreacy_N246}
\end{figure}

Figure \ref{fig:discreacy_N246} displays explicitly the discrepancies in the energies of the $k = 1, 2$ states, obtained from the IMSRG(2) and IMSRG(2*) methods relative to the exact solutions for systems with $N = 2, 4$, and $6$. It is shown clearly that  the discrepancies increase with  the pairing strength  $g$ for the two states. Consistent with the finding in Fig.~\ref{fig:Energies_different_N}, the discrepancies in the energies for $N=4$ are globally larger than those for $N=2, 6$.  It implies that the NO2B approximation is worse for the mid-shell nuclei than for the near closed-shell nuclei.  With the inclusion of the $\chi$ term in the IMSRG(2*),  these discrepancies are significantly reduced for the $k=1$ state, by $1-2$ orders of magnitude. However, for the $k=2$ state, the improvement appears generally in the regions with $g>1$.  

\begin{figure}[bt]
 \centering
\includegraphics[width=\columnwidth]{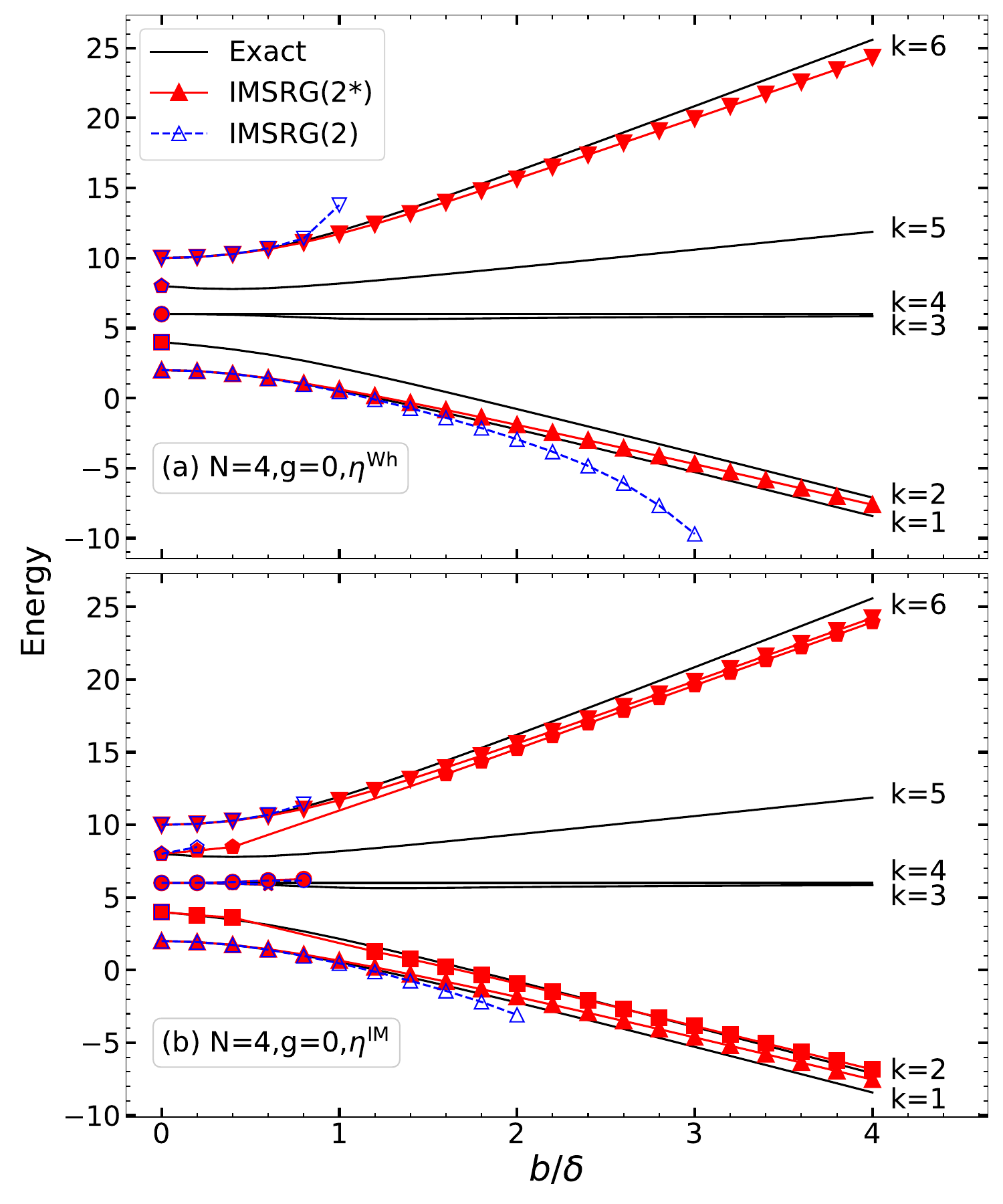}

\caption{(Color online) Comparison of energy levels calculated using the IMSRG(2) and IMSRG(2*) methods for the $N = 4$, $g = 0$ system as a function of the pair-breaking parameter $b/\delta$. The results of calculations using both (a) the White generator and (b) the imaginary-time generator are compared. The energy difference between results obtained with the two generators is typically within 1\%.
}
\label{fig:N4_levels_breaking_diff_generators}
\end{figure}

 \begin{figure}[bt]
\centering
\includegraphics[width=\columnwidth]{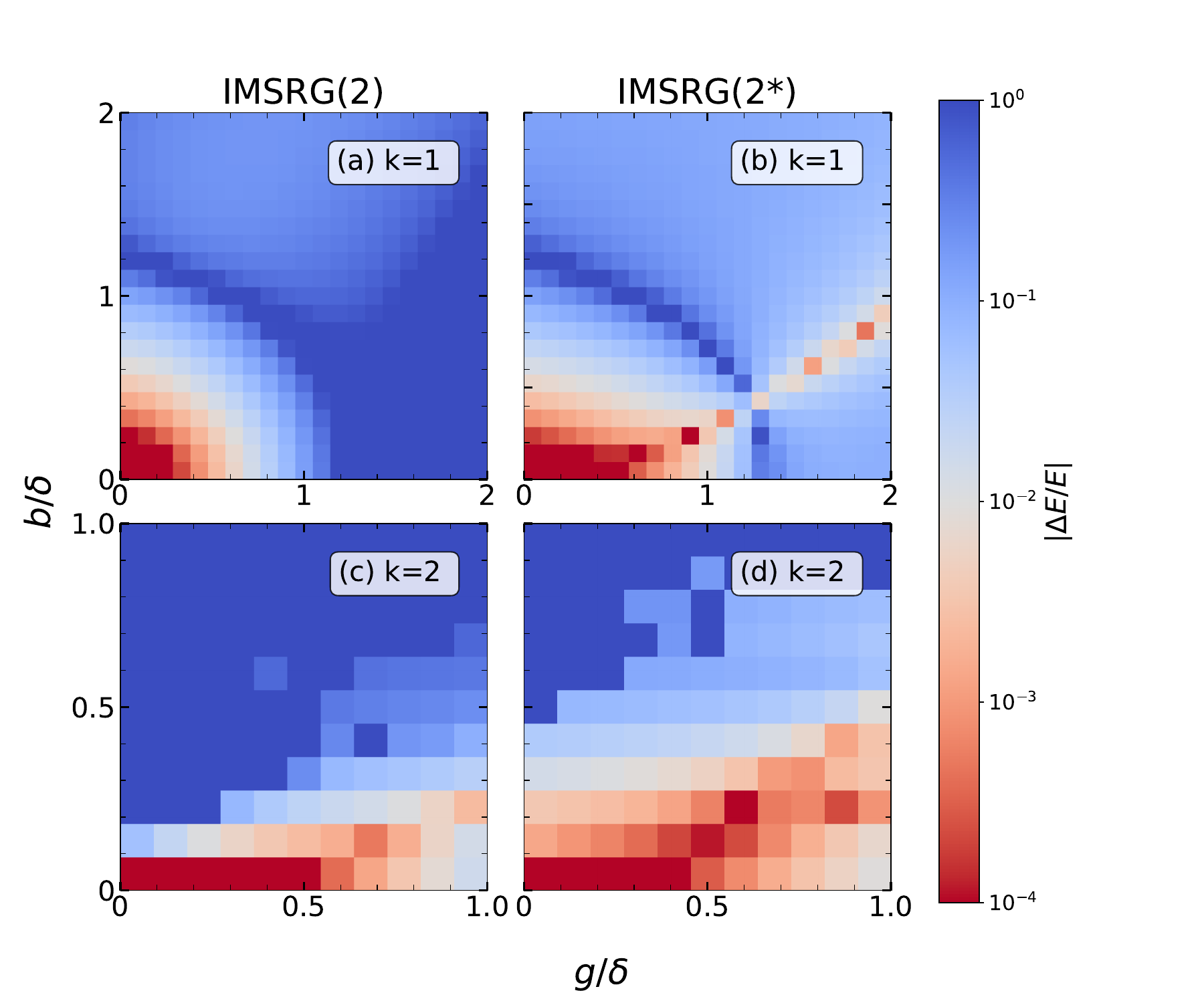}
\caption{(Color online)  Relative errors in the energies of the $k = 1, 2$ states, defined as $|E_{\mathrm{IMSRG}} - E_{\mathrm{exact}}| / E_{\mathrm{exact}}$, obtained using the IMSRG(2) and IMSRG(2*) methods as functions of the interaction strength $g$ and the pair-breaking parameter $b$ for the $N = 4$ system. See main text for details. }
 \label{fig:N4_Contour}
 \end{figure}

We benchmark the performance of IMSRG(2) and IMSRG(2*) for cases involving pair-breaking terms.  Figure \ref{fig:N4_levels_breaking_diff_generators} presents the energy levels of the $N=4, g=0$ system as a function of the pair-breaking strength parameter $b/\delta$ obtained from IMSRG(2) and IMSRG(2*) methods.  The states are labeled $k=1$ to $k=6$. It is shown that for some excited states, the White generator fails to converge. In contrast, the imaginary-time generator generally improves convergence, even though many instances remain where neither generator achieves a convergent solution. Moreover, one can see that IMSRG(2*) closely reproduces the exact results over the entire range, whereas IMSRG(2) shows larger deviations for certain states, particularly at higher $b/\delta$.   Overall, IMSRG(2*) demonstrates superior accuracy and convergence compared to IMSRG(2).

Figure~\ref{fig:N4_Contour} displays the relative errors in the energies of the $k = 1, 2$ states calculated with IMSRG(2) and IMSRG(2*) as functions of both $g$ and $b$ for the $N = 4$ system.   Again, the errors increase with the interaction strength, which can lead to convergence failure. In contrast to the standard IMSRG(2) approach, IMSRG(2*) not only provides higher accuracy but also demonstrates an extended convergence domain. Moreover, the figure shows that the convergence range for the $k = 2$ state is significantly narrower than that of the ground state, indicating that calculating excited-state energies with IMSRG(2) is particularly challenging.

 \begin{figure}[]
\centering
\includegraphics[width=\columnwidth]{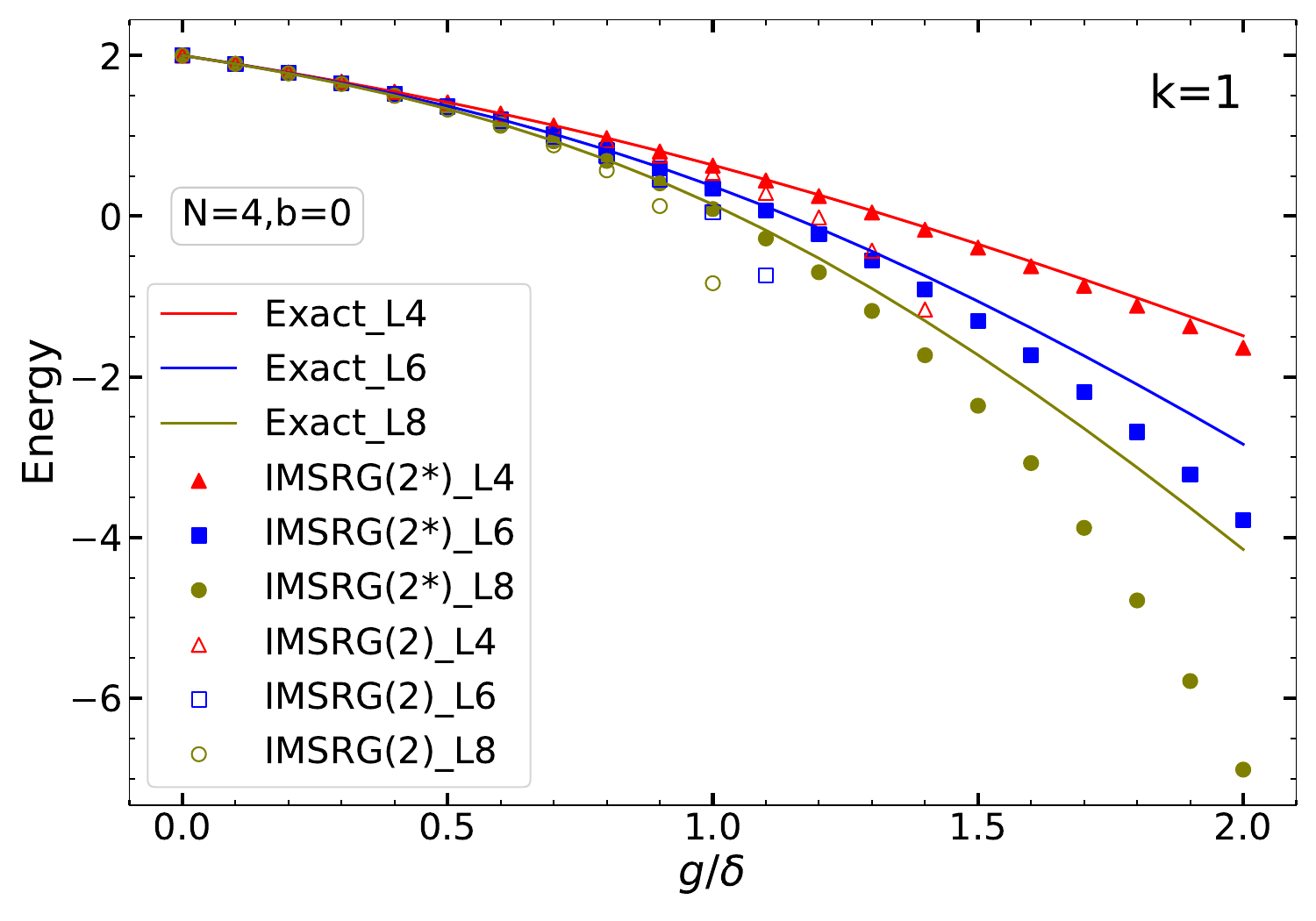}
\caption{(Color online) Ground-state energy ($k = 1$) as a function of the pairing strength $g/\delta$, calculated using the IMSRG(2) and IMSRG(2*) methods for the systems with $N = 4$ and $L = 4, 6,$ and $8$, respectively. Exact results (solid lines) are shown for comparison. The IMSRG(2*) results (solid markers) show significantly improved agreement with the exact solutions compared to IMSRG(2) (open markers), particularly in the strong-coupling regime and for larger model spaces.
 }
 \label{fig:N4_energy_L468}
 \end{figure}

 Finally, we benchmark the performance of the IMSRG(2) and IMSRG(2*) for the systems with $N = 4$ and single-particle model spaces of $L = 4, 6,$ and $8$, respectively. The ground-state energies  as a function of the pairing strength $g/\delta$, calculated using IMSRG(2), IMSRG(2*), and exact methods are displayed in Fig.~\ref{fig:N4_energy_L468}.   It is shown that IMSRG(2*) closely follows the exact results across all model spaces, while IMSRG(2) significantly deviates, especially at stronger interactions and larger model spaces. This indicates that IMSRG(2*) offers improved accuracy and robustness over IMSRG(2), particularly in strongly correlated systems and larger Hilbert spaces, making it a more reliable approach for realistic nuclear structure calculations.

\section{Summary}
\label{section:summary}
 
In this work, we have benchmarked the IMSRG(2) and IMSRG(2*) methods using a pairing-plus-particle-hole model for systems with varying particle numbers $N$, interaction strengths $(g, b)$, and single-particle model spaces $L$. The results demonstrate that IMSRG(2*) consistently yields a more accurate description of both ground and excited states compared to IMSRG(2), especially in the strongly interacting regime. These findings underscore the potential of IMSRG(2*) to overcome the large discrepancies and divergence issues that limit the applicability of standard IMSRG(2) in strongly correlated systems and extended Hilbert spaces. 

It is worth noting that despite these improvements, accurately describing systems with many valence particles and strong interactions, typical of mid-shell nuclei, remains a challenge for the IMSRG(2*). Previous studies have shown that employing multi-reference states can significantly enhance the treatment of collective correlations in such systems~\cite{Gebrerufael:2017PRL,Yao:2020PRL}. Building on this insight, work is currently in progress to extend the present study to the MR-IMSRG(2*) framework based on particle-number-projected Hartree-Fock-Bogoliubov reference states and chiral nuclear interactions, aiming for realistic applications to open-shell nuclei.

\begin{acknowledgments} 
We thank H. Hergert and J. M. Yao for fruitful discussions. This work is supported in part by the National Natural Science Foundation of China (Grant No. 12375119). This project is supported by the National Science Foundation (NSF) FRHTP program under award No. PHY-2402275.
 
\end{acknowledgments}

  \bibliography{ref}

\begin{thebibliography}{16}%
\makeatletter
\providecommand \@ifxundefined [1]{%
 \@ifx{#1\undefined}
}%
\providecommand \@ifnum [1]{%
 \ifnum #1\expandafter \@firstoftwo
 \else \expandafter \@secondoftwo
 \fi
}%
\providecommand \@ifx [1]{%
 \ifx #1\expandafter \@firstoftwo
 \else \expandafter \@secondoftwo
 \fi
}%
\providecommand \natexlab [1]{#1}%
\providecommand \enquote  [1]{``#1''}%
\providecommand \bibnamefont  [1]{#1}%
\providecommand \bibfnamefont [1]{#1}%
\providecommand \citenamefont [1]{#1}%
\providecommand \href@noop [0]{\@secondoftwo}%
\providecommand \href [0]{\begingroup \@sanitize@url \@href}%
\providecommand \@href[1]{\@@startlink{#1}\@@href}%
\providecommand \@@href[1]{\endgroup#1\@@endlink}%
\providecommand \@sanitize@url [0]{\catcode `\\12\catcode `\$12\catcode `\&12\catcode `\#12\catcode `\^12\catcode `\_12\catcode `\%12\relax}%
\providecommand \@@startlink[1]{}%
\providecommand \@@endlink[0]{}%
\providecommand \url  [0]{\begingroup\@sanitize@url \@url }%
\providecommand \@url [1]{\endgroup\@href {#1}{\urlprefix }}%
\providecommand \urlprefix  [0]{URL }%
\providecommand \Eprint [0]{\href }%
\providecommand \doibase [0]{http://dx.doi.org/}%
\providecommand \selectlanguage [0]{\@gobble}%
\providecommand \bibinfo  [0]{\@secondoftwo}%
\providecommand \bibfield  [0]{\@secondoftwo}%
\providecommand \translation [1]{[#1]}%
\providecommand \BibitemOpen [0]{}%
\providecommand \bibitemStop [0]{}%
\providecommand \bibitemNoStop [0]{.\EOS\space}%
\providecommand \EOS [0]{\spacefactor3000\relax}%
\providecommand \BibitemShut  [1]{\csname bibitem#1\endcsname}%
\let\auto@bib@innerbib\@empty
\bibitem [{\citenamefont {Hergert}(2020)}]{Hergert:2020abreview}%
  \BibitemOpen
  \bibfield  {author} {\bibinfo {author} {\bibfnamefont {H.}~\bibnamefont {Hergert}},\ }\href {\doibase 10.3389/fphy.2020.00379} {\bibfield  {journal} {\bibinfo  {journal} {Front. in Phys.}\ }\textbf {\bibinfo {volume} {8}},\ \bibinfo {pages} {379} (\bibinfo {year} {2020})},\ \Eprint {http://arxiv.org/abs/2008.05061} {arXiv:2008.05061 [nucl-th]} \BibitemShut {NoStop}%
\bibitem [{\citenamefont {Tsukiyama}\ \emph {et~al.}(2011)\citenamefont {Tsukiyama}, \citenamefont {Bogner},\ and\ \citenamefont {Schwenk}}]{Tsukiyama:2011PRL}%
  \BibitemOpen
  \bibfield  {author} {\bibinfo {author} {\bibfnamefont {K.}~\bibnamefont {Tsukiyama}}, \bibinfo {author} {\bibfnamefont {S.~K.}\ \bibnamefont {Bogner}}, \ and\ \bibinfo {author} {\bibfnamefont {A.}~\bibnamefont {Schwenk}},\ }\href {\doibase 10.1103/PhysRevLett.106.222502} {\bibfield  {journal} {\bibinfo  {journal} {Phys. Rev. Lett.}\ }\textbf {\bibinfo {volume} {106}},\ \bibinfo {pages} {222502} (\bibinfo {year} {2011})},\ \Eprint {http://arxiv.org/abs/1006.3639} {arXiv:1006.3639 [nucl-th]} \BibitemShut {NoStop}%
\bibitem [{\citenamefont {Stroberg}\ \emph {et~al.}(2021)\citenamefont {Stroberg}, \citenamefont {Holt}, \citenamefont {Schwenk},\ and\ \citenamefont {Simonis}}]{Stroberg:2021PRL}%
  \BibitemOpen
  \bibfield  {author} {\bibinfo {author} {\bibfnamefont {S.~R.}\ \bibnamefont {Stroberg}}, \bibinfo {author} {\bibfnamefont {J.~D.}\ \bibnamefont {Holt}}, \bibinfo {author} {\bibfnamefont {A.}~\bibnamefont {Schwenk}}, \ and\ \bibinfo {author} {\bibfnamefont {J.}~\bibnamefont {Simonis}},\ }\href {\doibase 10.1103/PhysRevLett.126.022501} {\bibfield  {journal} {\bibinfo  {journal} {Phys. Rev. Lett.}\ }\textbf {\bibinfo {volume} {126}},\ \bibinfo {pages} {022501} (\bibinfo {year} {2021})},\ \Eprint {http://arxiv.org/abs/1905.10475} {arXiv:1905.10475 [nucl-th]} \BibitemShut {NoStop}%
\bibitem [{\citenamefont {Hu}\ \emph {et~al.}(2022)\citenamefont {Hu} \emph {et~al.}}]{Hu:2022NP}%
  \BibitemOpen
  \bibfield  {author} {\bibinfo {author} {\bibfnamefont {B.}~\bibnamefont {Hu}} \emph {et~al.},\ }\href {\doibase 10.1038/s41567-023-02324-9} {\bibfield  {journal} {\bibinfo  {journal} {Nature Phys.}\ }\textbf {\bibinfo {volume} {18}},\ \bibinfo {pages} {1196} (\bibinfo {year} {2022})},\ \Eprint {http://arxiv.org/abs/2112.01125} {arXiv:2112.01125 [nucl-th]} \BibitemShut {NoStop}%
\bibitem [{\citenamefont {Yao}\ \emph {et~al.}(2020)\citenamefont {Yao}, \citenamefont {Bally}, \citenamefont {Engel}, \citenamefont {Wirth}, \citenamefont {Rodr\'{\i}guez},\ and\ \citenamefont {Hergert}}]{Yao:2020PRL}%
  \BibitemOpen
  \bibfield  {author} {\bibinfo {author} {\bibfnamefont {J.~M.}\ \bibnamefont {Yao}}, \bibinfo {author} {\bibfnamefont {B.}~\bibnamefont {Bally}}, \bibinfo {author} {\bibfnamefont {J.}~\bibnamefont {Engel}}, \bibinfo {author} {\bibfnamefont {R.}~\bibnamefont {Wirth}}, \bibinfo {author} {\bibfnamefont {T.~R.}\ \bibnamefont {Rodr\'{\i}guez}}, \ and\ \bibinfo {author} {\bibfnamefont {H.}~\bibnamefont {Hergert}},\ }\href {\doibase 10.1103/PhysRevLett.124.232501} {\bibfield  {journal} {\bibinfo  {journal} {Phys. Rev. Lett.}\ }\textbf {\bibinfo {volume} {124}},\ \bibinfo {pages} {232501} (\bibinfo {year} {2020})}\BibitemShut {NoStop}%
\bibitem [{\citenamefont {Belley}\ \emph {et~al.}(2024)\citenamefont {Belley}, \citenamefont {Yao}, \citenamefont {Bally}, \citenamefont {Pitcher}, \citenamefont {Engel}, \citenamefont {Hergert}, \citenamefont {Holt}, \citenamefont {Miyagi}, \citenamefont {Rodr\'{\i}guez}, \citenamefont {Romero}, \citenamefont {Stroberg},\ and\ \citenamefont {Zhang}}]{Belley:2024PRL}%
  \BibitemOpen
  \bibfield  {author} {\bibinfo {author} {\bibfnamefont {A.}~\bibnamefont {Belley}}, \bibinfo {author} {\bibfnamefont {J.~M.}\ \bibnamefont {Yao}}, \bibinfo {author} {\bibfnamefont {B.}~\bibnamefont {Bally}}, \bibinfo {author} {\bibfnamefont {J.}~\bibnamefont {Pitcher}}, \bibinfo {author} {\bibfnamefont {J.}~\bibnamefont {Engel}}, \bibinfo {author} {\bibfnamefont {H.}~\bibnamefont {Hergert}}, \bibinfo {author} {\bibfnamefont {J.~D.}\ \bibnamefont {Holt}}, \bibinfo {author} {\bibfnamefont {T.}~\bibnamefont {Miyagi}}, \bibinfo {author} {\bibfnamefont {T.~R.}\ \bibnamefont {Rodr\'{\i}guez}}, \bibinfo {author} {\bibfnamefont {A.~M.}\ \bibnamefont {Romero}}, \bibinfo {author} {\bibfnamefont {S.~R.}\ \bibnamefont {Stroberg}}, \ and\ \bibinfo {author} {\bibfnamefont {X.}~\bibnamefont {Zhang}},\ }\href {\doibase 10.1103/PhysRevLett.132.182502} {\bibfield  {journal} {\bibinfo  {journal} {Phys. Rev. Lett.}\ }\textbf {\bibinfo {volume} {132}},\ \bibinfo {pages} {182502} (\bibinfo {year} {2024})}\BibitemShut {NoStop}%
\bibitem [{\citenamefont {Hergert}\ \emph {et~al.}(2016)\citenamefont {Hergert}, \citenamefont {Bogner}, \citenamefont {Morris}, \citenamefont {Schwenk},\ and\ \citenamefont {Tsukiyama}}]{Hergert:2016PR}%
  \BibitemOpen
  \bibfield  {author} {\bibinfo {author} {\bibfnamefont {H.}~\bibnamefont {Hergert}}, \bibinfo {author} {\bibfnamefont {S.~K.}\ \bibnamefont {Bogner}}, \bibinfo {author} {\bibfnamefont {T.~D.}\ \bibnamefont {Morris}}, \bibinfo {author} {\bibfnamefont {A.}~\bibnamefont {Schwenk}}, \ and\ \bibinfo {author} {\bibfnamefont {K.}~\bibnamefont {Tsukiyama}},\ }\href {\doibase 10.1016/j.physrep.2015.12.007} {\bibfield  {journal} {\bibinfo  {journal} {Phys. Rept.}\ }\textbf {\bibinfo {volume} {621}},\ \bibinfo {pages} {165} (\bibinfo {year} {2016})},\ \Eprint {http://arxiv.org/abs/1512.06956} {arXiv:1512.06956 [nucl-th]} \BibitemShut {NoStop}%
\bibitem [{\citenamefont {Heinz}\ \emph {et~al.}(2021)\citenamefont {Heinz}, \citenamefont {Tichai}, \citenamefont {Hoppe}, \citenamefont {Hebeler},\ and\ \citenamefont {Schwenk}}]{Heinz:2021}%
  \BibitemOpen
  \bibfield  {author} {\bibinfo {author} {\bibfnamefont {M.}~\bibnamefont {Heinz}}, \bibinfo {author} {\bibfnamefont {A.}~\bibnamefont {Tichai}}, \bibinfo {author} {\bibfnamefont {J.}~\bibnamefont {Hoppe}}, \bibinfo {author} {\bibfnamefont {K.}~\bibnamefont {Hebeler}}, \ and\ \bibinfo {author} {\bibfnamefont {A.}~\bibnamefont {Schwenk}},\ }\href {\doibase 10.1103/PhysRevC.103.044318} {\bibfield  {journal} {\bibinfo  {journal} {Phys. Rev. C}\ }\textbf {\bibinfo {volume} {103}},\ \bibinfo {pages} {044318} (\bibinfo {year} {2021})},\ \Eprint {http://arxiv.org/abs/2102.11172} {arXiv:2102.11172 [nucl-th]} \BibitemShut {NoStop}%
\bibitem [{\citenamefont {Stroberg}\ \emph {et~al.}(2024)\citenamefont {Stroberg}, \citenamefont {Morris},\ and\ \citenamefont {He}}]{Stroberg:2024}%
  \BibitemOpen
  \bibfield  {author} {\bibinfo {author} {\bibfnamefont {S.~R.}\ \bibnamefont {Stroberg}}, \bibinfo {author} {\bibfnamefont {T.~D.}\ \bibnamefont {Morris}}, \ and\ \bibinfo {author} {\bibfnamefont {B.~C.}\ \bibnamefont {He}},\ }\href {\doibase 10.1103/PhysRevC.110.044316} {\bibfield  {journal} {\bibinfo  {journal} {Phys. Rev. C}\ }\textbf {\bibinfo {volume} {110}},\ \bibinfo {pages} {044316} (\bibinfo {year} {2024})}\BibitemShut {NoStop}%
\bibitem [{\citenamefont {He}\ and\ \citenamefont {Stroberg}(2024)}]{He:2024}%
  \BibitemOpen
  \bibfield  {author} {\bibinfo {author} {\bibfnamefont {B.~C.}\ \bibnamefont {He}}\ and\ \bibinfo {author} {\bibfnamefont {S.~R.}\ \bibnamefont {Stroberg}},\ }\href {\doibase 10.1103/PhysRevC.110.044317} {\bibfield  {journal} {\bibinfo  {journal} {Phys. Rev. C}\ }\textbf {\bibinfo {volume} {110}},\ \bibinfo {pages} {044317} (\bibinfo {year} {2024})},\ \Eprint {http://arxiv.org/abs/2405.19594} {arXiv:2405.19594 [nucl-th]} \BibitemShut {NoStop}%
\bibitem [{\citenamefont {Hjorth-Jensen}\ \emph {et~al.}(2017)\citenamefont {Hjorth-Jensen}, \citenamefont {Lombardo},\ and\ \citenamefont {van Kolck}}]{Hjorth-Jensen:2017}%
  \BibitemOpen
  \bibinfo {editor} {\bibfnamefont {M.}~\bibnamefont {Hjorth-Jensen}}, \bibinfo {editor} {\bibfnamefont {M.~P.}\ \bibnamefont {Lombardo}}, \ and\ \bibinfo {editor} {\bibfnamefont {U.}~\bibnamefont {van Kolck}},\ eds.,\ \href {\doibase 10.1007/978-3-319-53336-0} {\emph {\bibinfo {title} {{An Advanced Course in Computational Nuclear Physics}}}},\ Vol.\ \bibinfo {volume} {936}\ (\bibinfo  {publisher} {Springer},\ \bibinfo {year} {2017})\BibitemShut {NoStop}%
\bibitem [{\citenamefont {Davison}(2023)}]{Davison:2023}%
  \BibitemOpen
  \bibfield  {author} {\bibinfo {author} {\bibfnamefont {J.}~\bibnamefont {Davison}},\ }\emph {\bibinfo {title} {{Theoretical and computational improvements to the in-medium similarity renormalization group}}},\ \href {\doibase 10.25335/566c-a253} {Ph.D. thesis},\ \bibinfo  {school} {Michigan State U.} (\bibinfo {year} {2023})\BibitemShut {NoStop}%
\bibitem [{\citenamefont {Morris}\ \emph {et~al.}(2015)\citenamefont {Morris}, \citenamefont {Parzuchowski},\ and\ \citenamefont {Bogner}}]{Morris:2015}%
  \BibitemOpen
  \bibfield  {author} {\bibinfo {author} {\bibfnamefont {T.~D.}\ \bibnamefont {Morris}}, \bibinfo {author} {\bibfnamefont {N.}~\bibnamefont {Parzuchowski}}, \ and\ \bibinfo {author} {\bibfnamefont {S.~K.}\ \bibnamefont {Bogner}},\ }\href {\doibase 10.1103/PhysRevC.92.034331} {\bibfield  {journal} {\bibinfo  {journal} {Phys. Rev. C}\ }\textbf {\bibinfo {volume} {92}},\ \bibinfo {pages} {034331} (\bibinfo {year} {2015})},\ \Eprint {http://arxiv.org/abs/1507.06725} {arXiv:1507.06725 [nucl-th]} \BibitemShut {NoStop}%
\bibitem [{\citenamefont {White}(2002)}]{White2002}%
  \BibitemOpen
  \bibfield  {author} {\bibinfo {author} {\bibfnamefont {S.~R.}\ \bibnamefont {White}},\ }\href {\doibase 10.1063/1.1508370} {\bibfield  {journal} {\bibinfo  {journal} {The Journal of Chemical Physics}\ }\textbf {\bibinfo {volume} {117}},\ \bibinfo {pages} {7472} (\bibinfo {year} {2002})}\BibitemShut {NoStop}%
\bibitem [{\citenamefont {Hergert}(2017)}]{Hergert:2017PS}%
  \BibitemOpen
  \bibfield  {author} {\bibinfo {author} {\bibfnamefont {H.}~\bibnamefont {Hergert}},\ }\href {\doibase 10.1088/1402-4896/92/2/023002} {\bibfield  {journal} {\bibinfo  {journal} {Phys. Scripta}\ }\textbf {\bibinfo {volume} {92}},\ \bibinfo {pages} {023002} (\bibinfo {year} {2017})},\ \Eprint {http://arxiv.org/abs/1607.06882} {arXiv:1607.06882 [nucl-th]} \BibitemShut {NoStop}%
\bibitem [{\citenamefont {Gebrerufael}\ \emph {et~al.}(2017)\citenamefont {Gebrerufael}, \citenamefont {Vobig}, \citenamefont {Hergert},\ and\ \citenamefont {Roth}}]{Gebrerufael:2017PRL}%
  \BibitemOpen
  \bibfield  {author} {\bibinfo {author} {\bibfnamefont {E.}~\bibnamefont {Gebrerufael}}, \bibinfo {author} {\bibfnamefont {K.}~\bibnamefont {Vobig}}, \bibinfo {author} {\bibfnamefont {H.}~\bibnamefont {Hergert}}, \ and\ \bibinfo {author} {\bibfnamefont {R.}~\bibnamefont {Roth}},\ }\href {\doibase 10.1103/PhysRevLett.118.152503} {\bibfield  {journal} {\bibinfo  {journal} {Phys. Rev. Lett.}\ }\textbf {\bibinfo {volume} {118}},\ \bibinfo {pages} {152503} (\bibinfo {year} {2017})},\ \Eprint {http://arxiv.org/abs/1610.05254} {arXiv:1610.05254 [nucl-th]} \BibitemShut {NoStop}%
\end{thebibliography}%

\end{document}